\documentclass[12pt]{article}
\usepackage[cp1251]{inputenc}
\usepackage{epsfig,graphicx}
\begin{document}
\begin{center}

\textbf{\Large Modification of Coulomb law and energy
levels of hydrogen atom in superstrong magnetic field}\\[0.5cm]
 M.~I.~Vysotsky\\[0.5cm]

\begin{tabular}{l}
\textit{\indent ITEP, 117218 Moscow, Russia}\\
\textit{\indent e-mail: vysotsky@itep.ru}
\end{tabular}
\end{center}

\begin{abstract}
 The screening of a Coulomb potential by superstrong
magnetic field is studied. Its influence on the spectrum of a
hydrogen atom is determined.
\end{abstract}

\begin{tabular}{l}
\hbox{PACS: 31.30.jf}\hfill \\
\end{tabular}

\newpage

I will discuss recently solved Quantum Mechanical - Quantum
Electrodynamical problem in this lectures. It was solved
numerically in papers \cite{1, 2}, and then analytical solution
was found in papers \cite{3, 4}.

We will use convenient in atomic physics Gauss units: $e^2 =
\alpha = 1/137$. Magnetic fields $B > m_e^2 e^3$ we will call
strong while $B > m_e^2/e^3$ we will call superstrong. Important
quantity in the problem under consideration is Landau radius $a_H
= 1/\sqrt{eB}$ called magnetic length in condensed matter physics.

Let us consider hydrogen atom in external homogeneous magnetic
field $B$. At strong $B$ Bohr radius $a_B$ is larger than $a_H$,
so there are two time scales in the problem: fast motion in the
perpendicular to magnetic field plane and slow motion along the
magnetic field. That is why adiabatic approximation is applicable:
averaging over fast motion we get one-dimensional motion of
electron along the magnetic field in effective potential
\begin{equation}
U(z) \approx\frac{-e^2}{\sqrt{z^2 + a_H^2}} \label{1}
\end{equation}
The energy of a ground state can be estimated as
\begin{equation}
E_0 = -2m\left(\;\int\limits_{a_H}^{a_B} U(z) dz\right)^2 \sim
-me^4 \ln^2(B/m^2 e^3) \label{2}
\end{equation}
and it goes to minus infinity when $B$ goes to infinity.

We will see that radiative corrections qualitatively change this
result: ground state energy goes to finite value when $B$ goes to
infinity. This happens due to screening of Coulomb potential.

Since at strong $B$ reduction of the number of space dimensions
occurs and motion takes place in one space and one time dimensions
it is natural to begin systematic analysis from QED in $D=2$. At
tree level Coulomb potential is
\begin{equation}
\Phi(k) \equiv A_0(\bar k) = \frac{4\pi g}{\bar k^2} \; \; ,
\label{3}
\end{equation}
while taking into account loop insertions into photon propagator
we get:
\begin{equation}
\Phi(k) \equiv {\bf A}_0 = D_{00} + D_{00}\Pi_{00}D_{00} + ... =
-\frac{4\pi g}{k^2 + \Pi(k^2)} \;\; , \label{4}
\end{equation}
where photon polarization operator equals:
\begin{equation}
\Pi_{\mu\nu} \equiv\left(g_{\mu\nu}-\frac{k_\mu
k_\nu}{k^2}\right)\Pi(k^2) \;\; , \label{5}
\end{equation}
\begin{equation}
\Pi(k^2) = 4g^2\left[\frac{1}{\sqrt{t(1+t)}}\ln(\sqrt{1+t} +\sqrt
t) -1\right] \equiv -4g^2 P(t) \;\; , \label{6}
\end{equation}
$ t\equiv -k^2/4m^2$, and dimension of charge $g$ equals mass in
$D=2$.

Taking $k = (0, k_\parallel)$, $k^2 = -k_\parallel ^2$ for the
Coulomb potential in the coordinate representation we get:
\begin{equation}
{\bf\Phi}(z) = 4\pi g \int\limits^\infty_{-\infty} \frac{e^{i
k_\parallel z} dk_\parallel/2\pi}{k_\parallel^2 + 4g^2
P(k_\parallel^2 /4m^2)} \;\; , \label{7}
\end{equation}
and the potential energy for the charges $+g$ and $-g$ is finally:
$ V(z) = -g{\bf\Phi}(z)$.

In order to perform integration in (\ref{7}) we need simplified
expression for $P(t)$. Taking into account that asymptotics of
$P(t)$ are
\begin{equation}
P(t) = \left\{
\begin{array}{lcl}
\frac{2}{3} t & , & t\ll 1 \\
1 & , & t\gg 1
\end{array}
\right.
\label{8}
\end{equation}
let us take as an interpolating formula the following expression:
\begin{equation}
\overline{P}(t) = \frac{2t}{3+2t} \;\; .\label{9}
\end{equation}
The accuracy of this approximation is better than 10\%.

Substituting (\ref{9}) into (\ref{7}) we get:
\begin{eqnarray}
{\bf\Phi} & = & 4\pi g\int\limits^{\infty}_{-\infty} \frac{e^{i
k_\parallel z} d k_\parallel/2\pi}{k_\parallel^2 +
4g^2(k_\parallel^2/2m^2)/(3+k_\parallel^2/2m^2)} = \nonumber
\\
& = & \frac{4\pi g}{1+ 2g^2/3m^2}
\int\limits_{-\infty}^{\infty}\left[\frac{1}{k_\parallel^2} +
\frac{2g^2/3m^2}{k_\parallel^2 + 6m^2 + 4g^2}\right]
e^{ik_\parallel z} \frac{dk_\parallel}{2\pi} = \\
&=& \frac{4\pi g}{1+ 2g^2/3m^2}\left[-\frac{1}{2}|z| +
\frac{g^2/3m^2}{\sqrt{6m^2 + 4g^2}} {\rm exp}(-\sqrt{6m^2
+4g^2}|z|)\right] \;\; . \nonumber
\end{eqnarray}

In the case of heavy fermions ($m\gg g$)  the potential is given
by the tree level expression; the corrections are suppressed as
$g^2/m^2$.

In the case of light fermions ($m \ll g$):

\begin{equation}
{\bf\Phi}(z)\left|
\begin{array}{l}
~~  \\
m \ll g
\end{array}
\right. = \left\{
\begin{array}{lcl}
\pi e^{-2g|z|} & , & z \ll \frac{1}{g} \ln\left(\frac{g}{m}\right) \\
-2\pi g\left(\frac{3m^2}{2g^2}\right)|z| & , & z \gg \frac{1}{g}
\ln\left(\frac{g}{m}\right) \;\; .
\end{array}
\right. \label{11}
\end{equation}

For $m=0$ we have Schwinger model -- the first gauge invariant
theory with a massive vector boson. Light fermions make a
continuous transition from $m>g$ to $m=0$ case. The next two
figures correspond to $g=0.5$, $m=0.1$. The expression for $\bar
V$ contains $\bar P$.

\includegraphics[width=.8\textwidth]{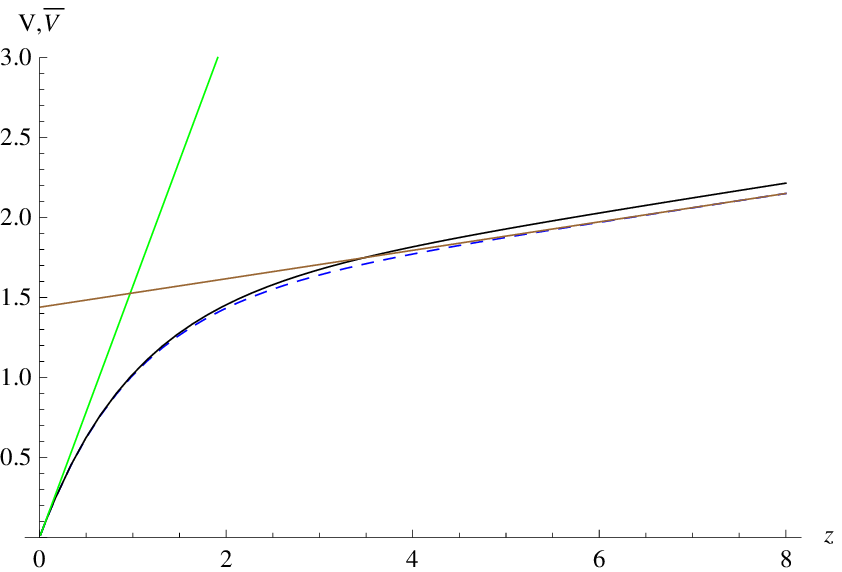}

Fig.1: Potential energy of the charges $+g$ and $-g$ in $D=2$. The
solid curve corresponds to $P$; the dashed curve corresponds to
$\bar P$.

\includegraphics[width=.8\textwidth]{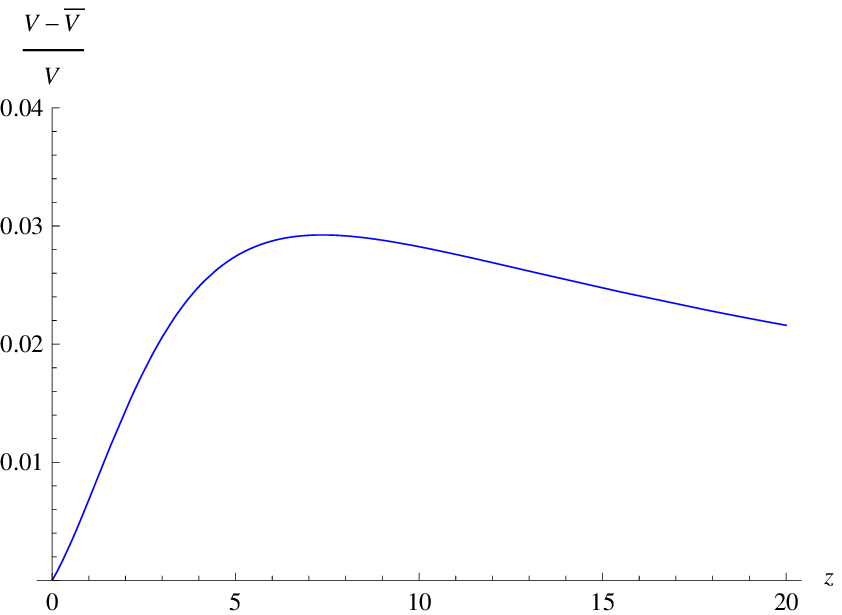}

Fig. 2: Relative difference of potential energies calculated with
the exact and interpolating formulae for the polarization operator
for $g=0.5$, $m=0.1$.
\bigskip

To find the modification of Coulomb potential in $D=4$ we need an
expression for $\Pi$ in strong $B$.

One starts from electron propagator $G$ in strong $B$. Solutions
of Dirac equation in homogenious constant in time $B$ are known,
so one can write spectral representation of electron Green
function. Denominators contain $k^2-m^2-2neB$, and for $B>>m^2/e$
and $k_\parallel^2<<eB$ in sum over levels 
lowest Landau level (LLL, $n=0$) dominates.
In coordinate representation transverse part of LLL wave function
is: $\Psi\sim exp((-x^2-y^2)eB)$ which in momentum representation
gives $\Psi\sim exp((-k_x^2-k_y^2)/eB)$ (gauge in which $\vec{A} =
1/2 [\vec{B} \times \vec{r}]$ is used).

Substituting electron Green functions into polarization operator
we get: \begin{eqnarray} \Pi_{\mu\nu} & \sim & e^2 eB
\int\frac{dq_xdq_y}{eB}exp(-\frac{q_x^2+q_y^2}{eB})\times \nonumber \\
&\times & exp(-\frac{(q+k)_x^2+(q+k)_y^2}{eB})dq_0dq_z\gamma_{\mu}
\frac{1}{\hat q_{0,z}-m}\gamma_{\nu}  \frac{1}{\hat q_{0,z}+\hat
k_{0,z}-m} = \nonumber \\
& = & e^3 B \times exp(-\frac{k^2_\bot}{2eB}) \times
\Pi_{\mu\nu}^{(2)}(k_\parallel\equiv
 k_z);   \label{12}
 \end{eqnarray}

\begin{equation}
{\bf\Phi} =\frac{4\pi e}{(k_\parallel^2 +
k_\bot^2)\left(1-\frac{\alpha}{3\pi}\ln\left(\frac{eB}{m^2}\right)\right)
+ \frac{2 e^3 B}{\pi} {\rm exp}\left(-\frac{k_\bot^2}{2eB}\right)
P\left(\frac{k_\parallel^2}{4m^2}\right)} \; . \label{13}
\end{equation}
\begin{equation}
{\bf\Phi}(z) = 4\pi e \int\frac{e^{ik_\parallel z} d k_\parallel
d^2 k_\bot/(2\pi)^3}{k_\parallel^2 + k_\bot^2 + \frac{2 e^3B}{\pi}
{\rm
exp}(-k_\bot^2/(2eB))(k_\parallel^2/2m_e^2)/(3+k_\parallel^2/2m_e^2)}
\;\; , \label{14}
\end{equation}

\begin{equation}
{\bf\Phi}(z) = \frac{e}{|z|}\left[ 1-e^{-\sqrt{6m_e^2}|z|} +
e^{-\sqrt{(2/\pi) e^3 B + 6m_e^2}|z|}\right] \;\; . \label{15}
\end{equation}
For magnetic fields $B \ll 3\pi m^2/e^3$ the potential is
Coulomb up to small power suppressed terms:
\begin{equation}
{\bf\Phi}(z)\left| \begin{array}{l}
~~  \\
e^3 B \ll m_e^2
\end{array}
\right. = \frac{e}{|z|}\left[ 1+ O\left(\frac{e^3
B}{m_e^2}\right)\right] \label{16}
\end{equation}
in full accordance with the $D=2$ case, where $g^2$ plays the role
of $e^3 B$.

In the opposite case of superstrong magnetic fields $B\gg 3\pi
m_e^2/e^3$ we get:
\begin{equation}
{\bf\Phi}(z) = \left\{
\begin{array}{lll}
\frac{e}{|z|} e^{(-\sqrt{(2/\pi) e^3 B}|z|)}  ,
\frac{1}{\sqrt{(2/\pi) e^3 B}}\ln\left(\sqrt{\frac{e^3 B}{3\pi
m_e^2}}\right)>|z|>\frac{1}{\sqrt{e B}}\\
\frac{e}{|z|}(1- e^{(-\sqrt{6m_e^2}|z|)})  ,  \frac{1}{m} > |z| >
\frac{1}{\sqrt{(2/\pi)e^3 B}}\ln\left(\sqrt{\frac{e^3 B}{3\pi
m_e^2}}\right) \\
\frac{e}{|z|}\;\; , \;\;\;\;\;\;\;\;\;\;\;\;\;\;\;\;\;\;\;\;\;\;\;
  |z| > \frac{1}{m}
\end{array}
\right. \;\; , \label{17}
\end{equation}

\begin{equation}
 V(z) = - e{\bf\Phi}(z)
\label{18}
\end{equation}

\begin{center}

\includegraphics[width=.8\textwidth]{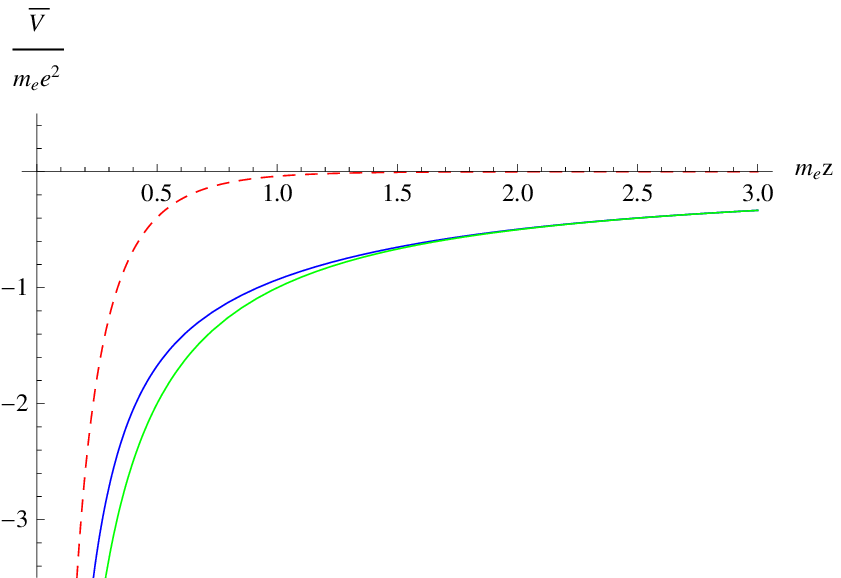}

Fig. 3: Modified Coulomb potential at $B=10^{17}$G (blue) and its
long distance (green) and short distance (red) asympotics.

\bigskip

\includegraphics[width=.8\textwidth]{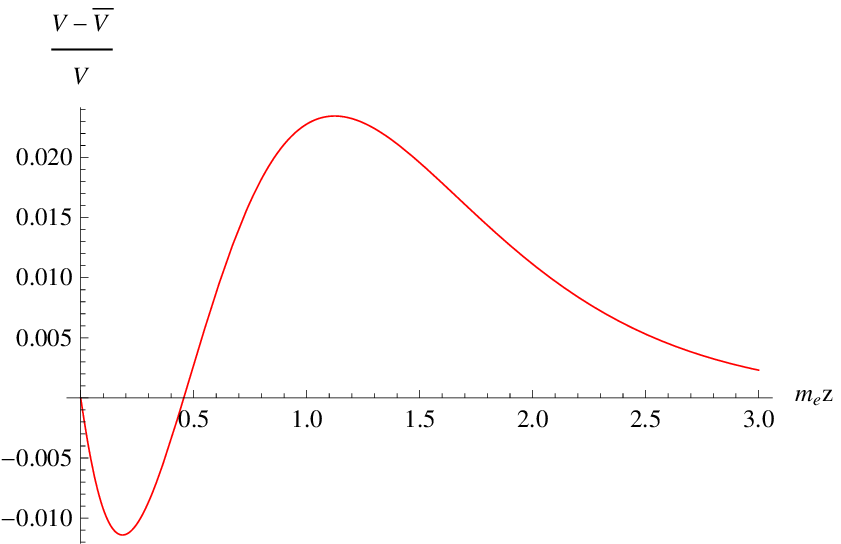}

Fig.4: Relative accuracy of analytical formula for modified
Coulomb potential at $B=10^{17} $G.

\end{center}

\bigskip

Spectrum of Dirac equation in constant in space and time magnetic
field is well known:
\begin{equation}
\varepsilon_n^2 = m_e^2 + p_z^2 + (2n + 1 + \sigma_z) eB \;\; ,
\label{19}
\end{equation}
$n= 0, 1, 2, 3,...; \;\; \sigma_z = \pm 1$. For $B > B_{cr} =
m_e^2/e$ the electrons are relativistic with only one exception:
electrons from lowest Landau level ($n=0, \;\; \sigma_z =
-1$) can be nonrelativistic.

In what follows we will study the spectrum of electrons from LLL
in the Coulomb field of the proton modified by the superstrong
$B$.

Spectrum of Schr\"{o}dinger equation in cylindrical coordinates
$(\bar\rho, z)$ in the gauge, where $\bar A = \frac{1}{2}[\bar B
\bar r]$ is:
\begin{equation}
E_{p_z n_\rho m \sigma_z} = \left(n_\rho + \frac{|m| +m+1+
\sigma_z}{2}\right)\frac{eB}{m_e} + \frac{p_z^2}{2m_e} \;\; ,
\label{20}
\end{equation}
LLL corresponds to $n_\rho = 0, \; \sigma_z = -1, \; m= 0, -1,
-2,...$.

A wave function factorizes on those describing free motion along a
magnetic field with momentum $p_z$ and those describing motion in
perpendicular to magnetic field plane:
\begin{equation}
R_{0m}(\bar\rho) = \left[\pi(2a_H^2)^{1+|m|} (|m|!)\right]^{-1/2}
\rho^{|m|}e^{(im\varphi - \rho^2/(4a_H^2))} \;\; , \label{21}
\end{equation}

Now we should take into account electric potential of atomic
nuclei situated at $\bar\rho = z = 0$. For $a_H \ll a_B$ adiabatic
approximation is applicable and the wave function in the following
form should be looked for:
\begin{equation}
\Psi_{n 0 m -1} = R_{0m}(\bar\rho)\chi_n(z) \;\; , \label{22}
\end{equation}
where $\chi_n(z)$ is the solution of the Schr\"{o}dinger equation
for electron motion along a magnetic field:
\begin{equation}
\left[-\frac{1}{2m} \frac{d^2}{d z^2} + U_{eff}(z)\right]
\chi_n(z) = E_n \chi_n(z) \;\; . \label{23}
\end{equation}
Without screening the effective potential is given by the
following formula:
\begin{equation}
U_{eff} (z) = -e^2\int\frac{|R_{0m}(\rho)|^2}{\sqrt{\rho^2 +
z^2}}d^2 \rho \;\; , \label{24}
\end{equation}
 For $|z| \gg a_H$ the effective potential equals Coulomb:
\begin{equation}
U_{eff}(z) \left|
\begin{array}{l}
~~  \\
z \gg a_H
\end{array}
\right. = - \frac{e^2}{|z|} \;\; \label{25}
\end{equation}
and effective potential is regular at $z=0$:

\begin{equation}
U_{eff}(0)
 \sim - \frac{e^2}{|a_H|} \;\; .
\label{26}
\end{equation}

Since $U_{eff}(z) = U_{eff}(-z)$, the wave functions are odd or
even under reflection $z\to -z$; the ground states (for $m=0, -1,
-2, ...$) are described by even wave functions.

To calculate the ground state of hydrogen atom in the textbook
``Quantum Mechanics'' by L.D.Landau and E.M.Lifshitz 
the shallow-well
approximation is used:

\begin{equation}
E^{sw} = -2m_e\left[\int\limits_{a_H}^{a_B} U(z) dz\right]^2 =
-(m_e e^4/2)ln^2(B/(m_e^2e^3)) \label{27}
\end{equation}

Let us derive this formula. The starting point is one-dimensional
Schr\"{o}dinger equation:
\begin{equation} -\frac{1}{2\mu}
\frac{d^2}{dz^2} \chi(z) + U(z) \chi(z) = E_0 \chi(z) \;\;
\label{28}
\end{equation}
Neglecting $E_0$ in comparison with $U$ and integrating we get:
\begin{equation}
\chi^\prime(a) = 2\mu\int\limits_0^a U(x)\chi(x) dx \;\; ,
\label{29}
\end{equation}
where we assume $U(x) = U(-x)$, that is why $\chi$ is even.

The next assumptions are: 1. the finite range of the potential
energy: $U(x) \neq 0$ for $a> x > -a$; 2. $\chi$ undergoes very
small variations inside the well. Since outside the well $\chi(x)
\sim e^{-\sqrt{2\mu |E_0|}\;x}$, we readily obtain:
\begin{equation}
|E_0| = 2\mu\left[\int\limits_0^a U(x) dx\right]^2 \;\; .
\label{30}
\end{equation}

For
\begin{equation}
\mu |U| a^2 \ll 1 \label{31}
\end{equation}
(condition for the potential to form a shallow well) we get that,
indeed, $|E_0| \ll |U|$ and that the variation of $\chi$ inside
the well is small, $\Delta \chi/\chi \sim \mu |U|a^2 \ll 1$.
Concerning the one-dimensional Coulomb potential, it satisfies
this condition  only for $a\ll 1/(m_e e^2)\equiv a_B$.

This explains why the accuracy of $log^2$ formula is very poor.

Much more accurate equation for atomic energies in strong magnetic
field was derived by B.M.Karnakov and V.S.Popov \cite{5}. 
It provides a
several percent accuracy for the energies of EVEN states for  $H >
10^3$ ($H \equiv  B/(m_e^2 e^3)$).

Main idea is to integrate Shr\"{o}dinger equation with effective
potential from $x=0$ till $x=z$, where $a_H<<z<<a_B$ and to equate
obtained expression for $\chi^\prime(z)/\chi(z)$ to the
logarithmic derivative of Whittaker function - the solution of
Shr\"{o}dinger equation with Coulomb potential, which
exponentially decreases at $z>>a_B$:
\begin{eqnarray}
&&2\ln\left(\frac{z}{a_H}\right) + \ln 2 - \psi(1+|m|) + O(a_H/z)
= \nonumber \\
&&  2\ln\left(\frac{z}{a_B}\right) + \lambda + 2\ln \lambda +
2\psi\left(1-\frac{1}{\lambda}\right) + 4\gamma + 2\ln 2 +
O(z/a_B) \;\; , \label{32}
\end{eqnarray}

\begin{equation}
E = -(m_e e^4/2)\lambda^2 \label{33}
\end{equation}

The energies of the ODD states are:
\begin{equation}
E_{\rm odd} = -\frac{m_e e^4}{2n^2} + O\left(\frac{m_e^2
e^3}{B}\right) \; , \;\; n = 1,2, ... \;\; . \label{34}
\end{equation}
So, for superstrong magnetic fields $B \sim m_e^2/e^3$ the
deviations of odd states energies
from the Balmer series are negligible.

When screening is taken into account an expression for effective
potential transforms into
\begin{equation}
\tilde U_{eff} (z) = -e^2\int  \frac{|R_{0m}(\vec{\rho})|^2}
{\sqrt{\rho^2 +z^2}} d^2\rho \left[1-e^{-\sqrt{6m_e^2}\;z} +
e^{-\sqrt{(2/\pi)e^3 B + 6m_e^2}\;z}\right]
 \;\;  \label{35}
\end{equation}

The original KP equation for LLL splitting by Coulomb potential
is:
\begin{equation}
\ln(H) = \lambda + 2\ln\lambda +
2\psi\left(1-\frac{1}{\lambda}\right) + \ln 2 + 4\gamma +
\psi(1+|m|) \;\; , \label{36}
\end{equation}
where 
$\psi(x)$ is the logarithmic derivative of the gamma function; it
has simple poles at $x=0,-1,-2,...$.

The modified KP equation, which takes screening into account looks
like:
\begin{equation}
\ln\left(\frac{H}{1+\displaystyle\frac{e^6}{3\pi}H}\right) =
\lambda + 2\ln\lambda + 2\psi\left(1-\frac{1}{\lambda}\right) +
\ln 2 + 4\gamma + \psi(1+|m|) \;\; , \label{37}
\end{equation}
$E = -(m_e e^4/2)\lambda^2$. In particular, for a ground state
$\lambda = -11.2$, $E_0 = -1.7$ keV.

\bigskip

\begin{center}

\includegraphics[width=.48\textwidth]{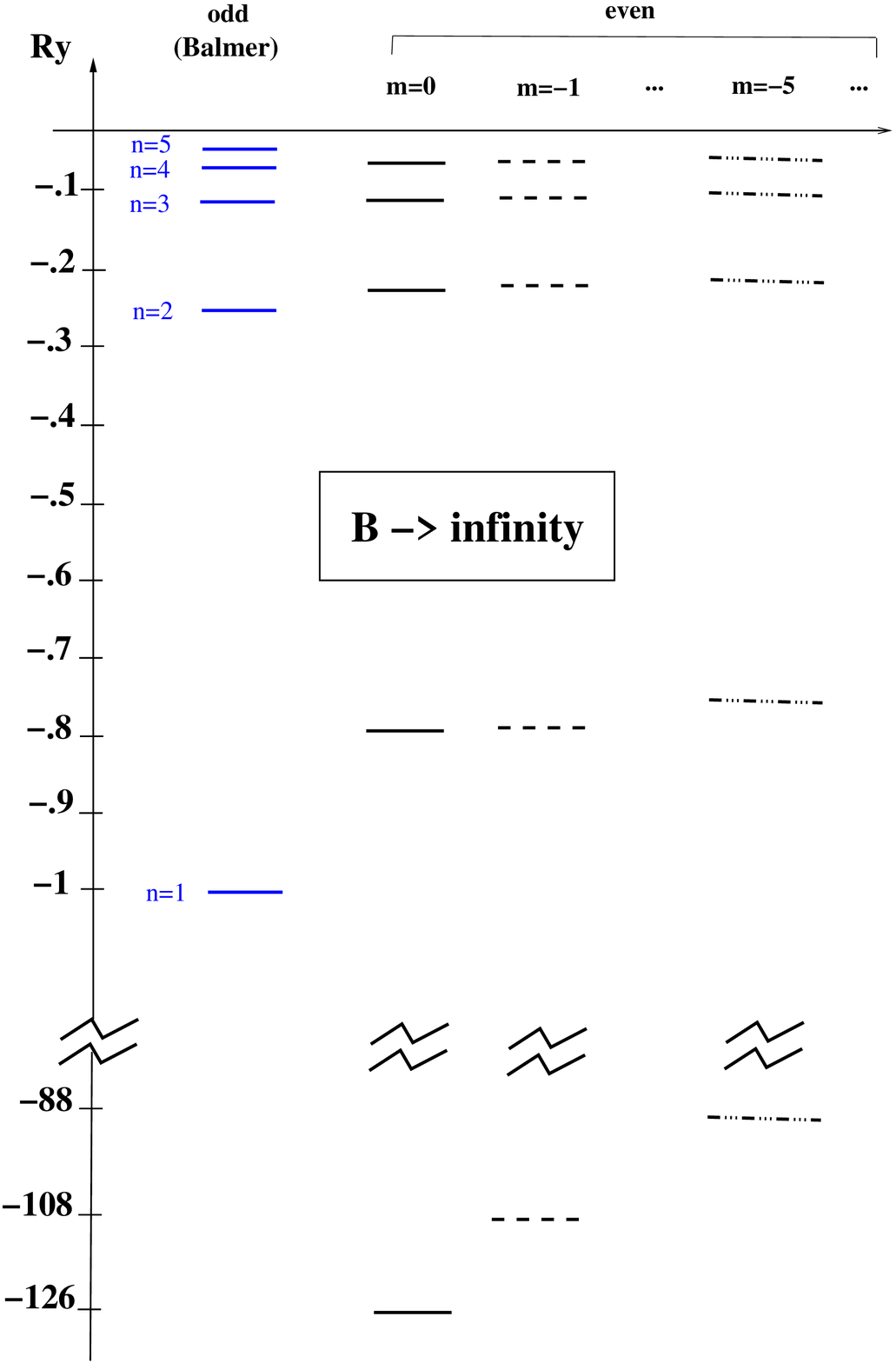}

\bigskip

Fig.5: Spectrum of hydrogen levels in the limit of infinite
magnetic field. Energies are given in rydberg units, $Ry \equiv
13.6 \; eV$.

\bigskip

\end{center}

In conclusion,

1. analytical expression  for charged particle electric potential
in $d=1$ is given; for $m<g$ screening take place at all
distances;

2. analytical expression  for charged particle electric potential
 $\Phi(z,\rho=0)$
at superstrong $B$ at $d=3$ is found; screening
take place at distances $|z|<1/m_e$;

3. an algebraic formula for the energy levels of a hydrogen atom
originating from the lowest Landau level in superstrong $B$ has
been obtained.

\bigskip
I am grateful to the organizers of Baikal Summer School for 
their hospitality. I was supported by the grant RFBR 11-02-00441.

\end{document}